**Population pharmacokinetics of levobupivacaine during a transversus abdominis plane block in children.**

*Marc Vincent [1] Olivier Mathieu [2] Patrick Nolain [3] Cecilia Menacé [4] and Sonia Khier [1,5]*

[1] PharmD, Department of Pharmacokinetics, School of Pharmacy, Montpellier University, Montpellier, France

[2] PharmD/PhD, Department of Toxicology and Target drug monitoring, Montpellier University Hospital (CHU Lapeyronie), Montpellier, France

[3] PharmD, Department of Pharmacometrics, Sanofi, Montpellier, France

[4] MD, Department of Anaesthesiology and Intensive Care, Montpellier University Hospital (CHU Lapeyronie), Montpellier, France

[5] PharmD/PhD, Institut Montpelliérain Alexander Grothendieck (IMAG), Probabilities and Statistics Department, CNRS UMR 5149, UMR 5149, Montpellier University

**Address for correspondence:**

Dr. Sonia Khier

ORCID ID https://orcid.org/0000-0001-6712-8461

ResearcherID: A-7317-2012

Associate Professor, Department of Pharmacokinetics - School of Pharmacy, Montpellier University

15 avenue Charles Flahault - Montpellier, 34000, France

sonia.khier@umontpellier.fr

+33 4 11 75 95 75

**Abstract**

**Background:** Levobupivacaine is commonly used in transversus abdominis plane block in paediatrics. However, dosing regimen is still empirical and is currently not relied on pharmacokinetic properties of levobupivacaine. We described pharmacokinetics of levobupivacaine during an ultrasound-guided transversus abdominis plane block in order to optimize dosing regimen, according to the between subject variability and the volume of levobupivacaine injected.

**Method:** The clinical trial (prospective, randomized, double-blind study protocol) was conducted in 40 children from 1 to 5 years old and planned for inguinal surgery. Each patient received 0.4 mg/kg of levobupivacaine with a volume of the local anaesthesia solution adjusted to 0.2 mL/kg of 0.2% or 0.4 mL/kg of 0.1% levobupivacaine. Blood samples were collected at 5, 15, 20, 25, 30, 45, 60 and 75 min following the block injection. The population pharmacokinetic analysis was performed with the NONMEM software.

**Results:** From the pharmacokinetic parameters obtained, median Cmax, tmax and area under the concentration *versus* time curve were respectively evaluated to 0.315 mg/L, 17 min and 41 mg/L. min. Between subject variability (BSV) of clearance is explained by the weight. The volume of the local anaesthesia injected influences bioavailability and so systemic concentration. At the dose regimen of 0.4 mg/kg, none of the infants presented toxicity signs, but for 13 patients transversus abdominis plane block failed. After analysis, BSV for absorption rate constant, distribution volume and clearance were 81%, 47% and 41% respectively. Residual unexplained variability was estimated to 14%.

**Conclusion:** Results suggest that, for a better efficiency in paediatric population, the dose of levobupivacaine should be greater than 0.4 mg/kg and the volume of solution injected must be high. Children's' weight should be considered to anticipate any risk of toxicity.

## 1. Introduction

Over the past years transversus abdominis plane (TAP) block, a loco regional anaesthesia technique, has become increasingly popular in abdominal surgery because it provides very effective pain relief both pre- and post-operatively thereby reducing postoperative opioid requirement. A local anaesthetic solution is injected into a neurovascular space, between the transverse and internal oblique muscle of the abdomen. The solution anesthetizes the nerves supplying the anterior abdominal wall (T9 to L1) [1,2], Figure X.

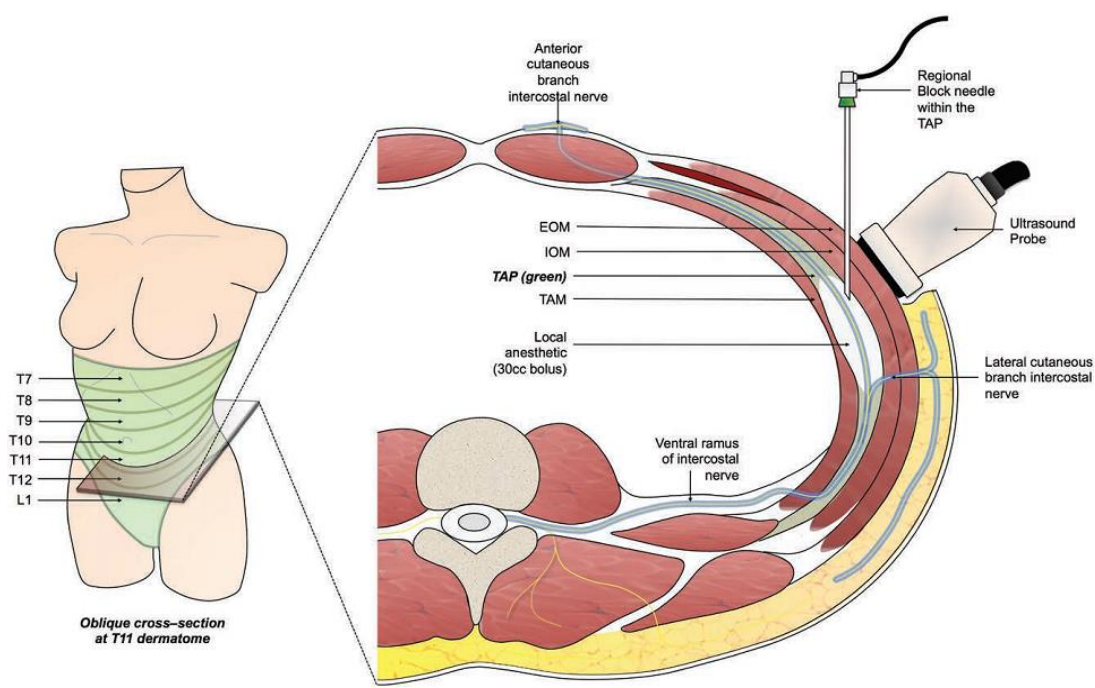

Fig. x Intraoperative transversus abdominis plane (*TAP*) blockade. Under ultrasound guidance, a regional block catheter is passed into the transversus abdominis plane at the anterior-axillary line, midway between the costal margin and iliac crest. An oblique cross-section at the level of T11 is shown (*right*). *EOM*, external oblique muscle; *IOM*, internal oblique muscle; *TAM*, transversus abdominis muscle. From Jablonka et al.

Levobupivacaine is commonly used as an anaesthetic in TAP block. The solution diffuses into the nerve, and causes a reversible block of the sodium channels. This prevents depolarisation and conduction of the nerve action potential. Levobupivacaine local concentration is correlated to its efficiency. But the solution is also absorbed in the systemic and may induce adverse effects (carditoxicity, neurotoxicity) [3]. Absorption from the site of administration is affected by the vascularity of the tissue. Protein binding is upper to 97%. Levobupivacaine is extensively metabolized with no unchanged levobupivacaine detected in urine or faeces. CYP3A4 and CYP1A2 mediate the metabolism of levobupivacaine to desbutyl-levobupivacaine and 3-hydroxy levobupivacaine, respectively The dosing regimen is still empirical in paediatric surgery and rarely based on pharmacokinetic data. In practice, an injected volume of 0.2 à 0.3 ml/kg of levobupivacaine (2.5 mg/ml) is used [4,5].

The aim of this work was therefore to describe pharmacokinetics of levobupivacaine and investigate possible sources of between subject variability in children by a population approach when performing an ultrasound-guided TAP block in ambulatory abdominal wall surgery. As the volume of anaesthetic injected could possibly influence its pharmacokinetics [6], the study was conducted according to two methods of administration: high or low volume of local anaesthetic injected.

## 2. Methods

This prospective, randomized, double-blind study protocol was approved by the institutional ethical committee (CPP Sud Mediterranean IV, 2013-A00618-37, Montpellier, France), and registered in Clinical Trials Database (ClinicalTrials.gov. NCT02064088). The written informed consent was obtained from the parents or legal guardians of all participating subjects before randomization.

### 2.1 Patients

Children from 1 to 5 years old (14 to 71 months) and planned for inguinal surgery were candidate to the study. Exclusion criteria included any contraindication to general anaesthesia or TAP block, amide local anaesthetic drug allergy, American Society of Anesthesiologists physical status 3 and 4, non-steroidal anti-inflammatory drug intolerance, known cardiac / renal / hepatic dysfunction, parental refusal, and concurrent participation to another clinical trial.

### 2.2 Study design and sampling protocol

Each patient received 0.4 mg/kg of levobupivacaine (Chirocaine®, Abbott, Rungis, France) [7] and were assigned to one of the two study groups using a computer generated random sequence. In group 1 (20 patients), the volume of the local anaesthesia solution was adjusted to 0.2 mL/kg of 0.2% levobupivacaine (low volume/high concentration). In group 2 (20 patients), the volume of the local anaesthesia solution was adjusted to 0.4 mL/kg of 0.1% levobupivacaine (high volume/low concentration). Blood samples were collected at 5, 15, 20, 25, 30, 45, 60 and 75 min following the block injection. To quantify the unbound fraction of levobupivacaine, orosomucoid ($\alpha$1-acid glycoprotein) plasma was immediately ultra-filtrated (cut-off 30 kDa, Ultracel®, Merck Millipore®, Tullagreen, Ireland) at room temperature and stored at -20°C until analysis. For plasma, the assay was done by simple protein precipitation in acetonitrile with added internal standard (etidocaine) and dilution in the mobile phase. The analysis of the prepared samples was performed by liquid chromatography combined with a hybrid triple quadrupole trap mass spectrometer (3200 Qtrap, ABSciex®, Les Ulis, France). Free levobupivacaine concentrations were analyzed by direct injection of the ultra-filtrated serum. According to a previously published method regarding ropivacaïne, the inter- and intra-day percentage coefficients of variation were below 8 % and 6%, respectively. The lowest limit of quantification (LLOQs) was 1 ng/ml [8].

**2.3 Pharmacokinetic analysis**

Pharmacokinetic analysis was performed with total concentrations of levobupivacaine, by non-linear mixed-effects modelling using NONMEM® version 7.4 (ICON Development Solutions, MD, USA). Based on a visual inspection of the individual pharmacokinetic profiles and a review of the literature, one- and two-compartment models were considered to describe the concentration-time data. Different kinetics of absorption (zero- or first- order) were investigated. The between-subject variability (BSV) for each model parameter was modelled using Equation 1 :

$$Pi = Ppop.\,e^{\eta i}, \eta \sim N(0, \omega^2)\ (1)$$

where Pi is the value of the parameter "(e.g., ka or absorption rate constant, CL or clearance, V or volume of distribution) for the $i_{th}$ patient, Ppop is the value of the population parameter and η is the random variable. Usual error models (additive, proportional and combined) were evaluated to describe residual unexplained variability during the screening step. The BSV is modelled in terms of random effect (η) variables. Each of these variables is assumed to have mean 0 and a variance denoted by $\omega^2$, which is estimated. Population parameters (typical values and BSV) were estimated using the FOCE-I algorithm (First Order Conditional Estimates with Interaction).

The model selection was performed on the basis of the minimization status (i.e., successful) and the examination of standard error of the parameter estimates (i.e., $\leq 50$ %). Models were compared with regards to the objective function value (OFV = -2. log-likelihood). The best model was selected based on the lowest OFV and the inspection of goodness of fits plots, i.e., observed plasma concentrations *versus* population predicted concentrations (PRED) plots, observed plasma concentrations *versus* individual predicted concentrations (IPRED) plots and conditional weighted residuals (CWRES) *versus* time, PRED or IPRED scatterplots.

Demographic characteristics (age, weight and gender) as well as the orosomucoide rate (Table 1) and the volume of the local anaesthesia solution injected (0.2 or 0.4 mL/kg) were evaluated as potential covariates. All potential covariates were added separately to the model (forward inclusion-backward elimination method) using different functions (linear, power, and exponential). All the functions were tested at each step of the building model. The significance of a parameter-covariate relationship was reflected by a decrease of OFV of at least 3.84 (p-value < 0.05, from a chi-squared distribution, with one degree of freedom).

| | N | % | Mean | Median | [Q1a-Q3b] |
|---|---|---|---|---|---|
| **Demographic and biological characteristics** | | | | | |
| Age (month) | 40 | | 42.8 | 43 | [28 - 54.7] |
| Weight (kg) | 40 | | 14.7 | 15 | [12 - 18] |
| Orosomucoide (g/L) | 37 | | 0.7 | 0.65 | [0.5 -0.8] |
| Gender | | | | | |
| Females | 12 | 30 | | | |
| Males | 28 | 70 | | | |
| | | | | | |
| **Individual exposure values** | | | | | |
| AUC | - | - | - | 41 | [20 - 73] |
| Cmax (mg/L) | - | - | - | 0.315 | [0.247 - 0.455] |
| Cu,max (µg/L) | - | - | - | 3.15 | - |
| Tmax (min) | - | - | - | 17 | [12 - 22] |

Table 1 Demographic and biological and individual exposures values. Cu,max is the unbound levobupivacaine concentration, estimated from the total concentration and unbound protein binding ((1-f) = 1%). The average value of unbound levobupivacaine was estimated to 1%. a: first quartile b: third quartile.

**2.4 Model evaluation**

The robustness of the model and the accuracy of parameters estimates were assessed using a bootstrap method [9]. The entire procedure was undertaken in an automated fashion. From the original dataset, 1 000 bootstrap sets were drawn with re-sampling. For each of the 1 000 bootstrap sets, the pharmacokinetic population parameters were estimated. The corresponding mean, median, standard deviation, 2.5 and 97.5th percentiles were calculated. Visual predictive checks (VPC) [10] were also performed to ensure that simulations based on model could reproduce the observed data. Perl-speaks-NONMEM Tools were used for both VPC and bootstrap evaluation [11].

## 2.5 Computation of individual exposures

Based on Empirical Bayes Estimates (EBE), single dose exposure values were calculated:

i)    area under the concentration *versus* time curve: $_0^\infty AUC = Dose / \left(\frac{Cl}{F}\right)$,

ii)   peak concentration $Cmax = \left(\frac{F \times Dose \times Ka}{V \times (Ka - Ke)}\right) \times \left(e^{-Ke.tmax} - e^{-Ka.tmax}\right)$,

iii)  time to reach Cmax : $tmax = ln\left(\frac{Ka}{Ke}\right) / (Ka - Ke)$.

## 2.6 TAP effectiveness evaluation

The cutaneous incision was allowed after a minimum of 15 minutes post TAP injection. Insufficient block (TAP block failure) was defined as an increase of either heart rate or mean arterial blood pressure ≥ 20% at the skin incision, compared to reference values prior block placement. In case of insufficient block, remifentanil infusion was used as backup.

## 2.7 Statistical analysis

Standard statistical tests (t-test, independent Wilcoxon or Fisher's Exact Test) were performed to compare the two groups "TAP Successful" and "TAP Failed". All the pharmacokinetic parameters and exposure values were compared between the two groups. Statistical analyses were performed using R version 3.5.1.

## 3. Results

### 3.1 Pharmacokinetic analysis

A total of 247 sample points from 40 subjects were included in the dataset. All the concentrations under the low limit of quantification (LLOQ = 0.05 mg/L) were considered as missing values (MDV=1).

The data were best described by a one-compartment model parameterized in terms of an absorption rate constant (ka, h-1) characterizing the first-order absorption process from the depot to the central compartment, an apparent distribution volume (V/F, L) and clearance (CL/F). The residual unexplained variability was described with a proportional error model (Equation 2):

$$C_i = C_{pred} + C_{pred} \times \sigma \quad (2)$$

where Ci is the concentration observed to the ith time and Cpred the concentration predicted by the model.

Weight significantly decreased the objective function and BSV (from 50 to 41%) of clearance and was integrated as covariate:

$$\left(\frac{Cl}{F}\right)i = \left(\frac{Cl}{F}\right)pop \, . \, (\frac{WT}{15})^{\,weight\,effect} \, . \, e^{\eta_{Cli}} \, . \, (3)$$

Where Cli is the clearance for the ith individual and Cl the value of population clearance.

Population pharmacokinetic parameters are presented in Table 2. Figure 1 demonstrates the quality of fit for pharmacokinetic data. Goodness-of-fit plots showed good agreement between predicted and observed concentrations, with no bias in residuals over time and across predicted concentration values. Individual concentration predictions are based on maximum values of a posteriori Bayesian estimates (posthoc option) while predicted typical (population) concentrations are based on population parameters. VPC are presented in Figure 2 and show a good predictive ability of the model.

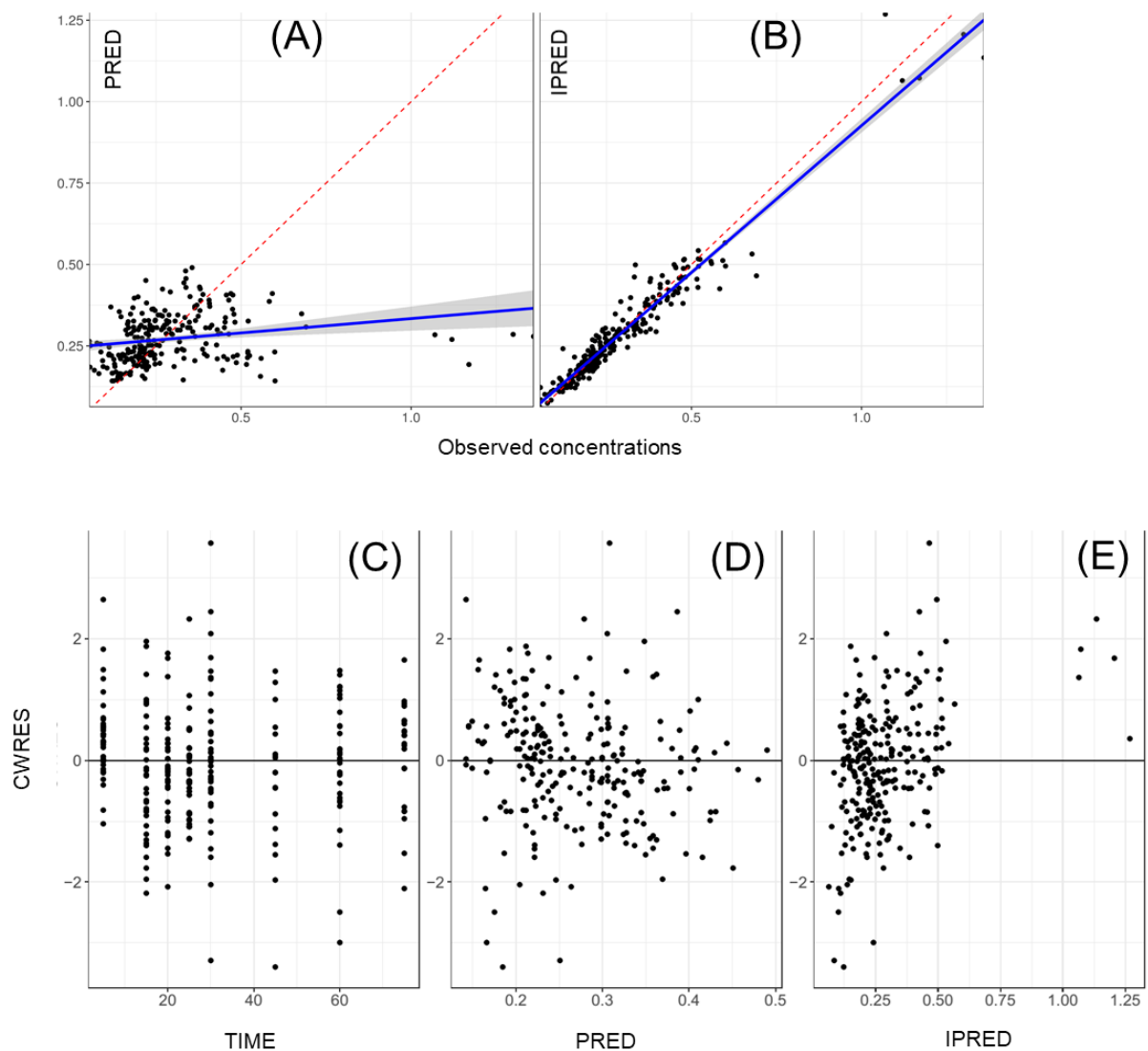

**Fig. 1** Commonly goodness of fit plots for the levobupivacaine model are used for evaluating model misspecification. Observed concentrations *versus*: (A) Population Predicted concentrations or PRED, (B) Individual Predicted concentrations or IPRED.  Conditional weighted residuals (CWRES) versus: (C) time, (D) PRED or (E) IPRED.

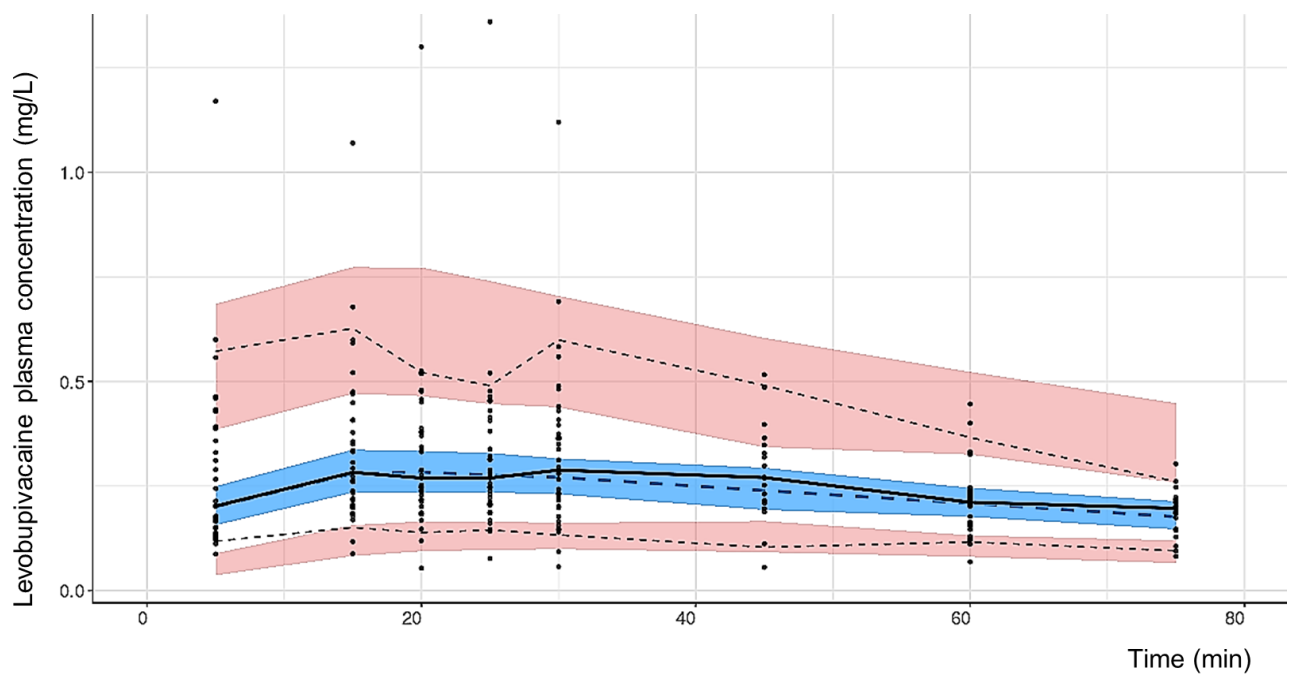

**Fig. 2** Visual predictive checks (VPC) plot generated from 1 000 simulations. Pink areas represent the 95 % confidence intervals (CI) of the 5th and 95th percentiles of the predictions. Upper dotted line: 95th percentile of the observations, lower dotted line: 5th percentile of the observations. Blue area represents the 95 % CI of the median predictions. Middle dotted line: median of predictions, middle solid line: median of observations. Black dots represent observed plasma concentrations.

The median maximum concentration was 0.315 mg/L at 17 minutes after the injection and the median total drug exposure across time was 41 mg/L. min. Mean protein binding estimate was 99 %. All the individual exposures values are presented in Table 1. Individual pharmacokinetic profiles were obtained and presented in Supplementary Data.

| Population pharmacokinetic parameters | Estimate | RSE (%) | [95% CI] | Shrinkage (%) | Bootstrap Median [2.5th Qu - 97.5th Qu] |
|---|---|---|---|---|---|
| Cl/F (L/min) | 0.15 | 14 | [0.10-0.19] | - | 0.14 [0.08 – 0.16] |
| Effect of weight on Cl/F | 0.87 | 44 | [0.08-1.65] | - | 1.16 [0.10 – 1.49] |
| V/F (L) | 14 | 12 | [11-18] | - | 15 [11 - 19] |
| Ka (/min) | 0.18 | 19 | [0.11-0.25] | - | 0.18 [0.14 – 0.23] |
| | | | | | |
| F (Small volume) | 1 FIX | - | - | - | - |
| F (Large volume) | 0.88 | 16 | [0.59-1.17] | - | 0.87 [0.63 – 1.15] |
| Between subject variability | Estimate (%, CV) | % RSE | [95% CI] | | |
| Cl/F | 41 | 45 | [10-56] | 12 | 43 [23 – 62] |
| V/F | 47 | 30 | [30-60] | 2.5 | 46 [29 – 61] |
| Ka | 81 | 25 | [57-98] | 14 | 80 [28 – 111] |
| Residual variability | 14 | 10 | [14-18] | 16.5 | 16 [13 – 19] |

Table 2 Population parameters obtained from the final model. CI: confidence interval, Qu: quartile, CV: coefficient of variation, RSE relative standard error (100% x SE/ Estimate), F bioavailability factor, Cl: clearance, V: volume of distribution of central compartment, ka first-order absorption rate constant.

### 3.2 Pharmacodynamics of levobupivacaine

Among the 40 patients, none of the 1 to 5 years-old children presented toxicity signs. An insufficient TAP block anaesthesia was obtained for 13 patients and a total anaesthesia was obtained for 27 patients. Pharmacokinetic parameters were not significantly different between the two groups: effective TAP block *versus* failed TAP block. PK profiles are presented in Supplementary Data. But we observed that gender and effectiveness of TAP block were dependent variables. (Fisher's Exact Test, p-value = 0.03).

## 4. Discussion

Levobupivacaine 0.25% can be administered with doses ranging up to 1 mL/kg [4,5,7,12–15].These recommendations are rarely based on pharmacokinetic data. Previous studies with ropivacaine showed an influence of the injected volume on the effectiveness of regional anaesthesia [6,16] while no study investigated the influence of the injected volume of levobupivacaine during a TAP block. The purpose of this work was to describe the pharmacokinetics of levobupivacaine in children and investigate possible sources of between subject variability including volumes of solution injected (small volume/high concentration or high volume/low concentration) during TAP block.

The absorption rate from TAP to systemic circulation is fast, with a mean absorption half-life = 6 min.

BSV on Ka has been estimated at 81 %. The forward inclusion-backward elimination method did not identify a covariate for Ka, and particularly the injected volume considered as categorical covariate (high/low injected volume). So, from our results, the injected volume does not influence Ka. Bioavailability factor (F) for "high volume injected" group is comparable to "small volume injected" group (F=0.88, [0.59-1.17]). It is important to note that whatever the group, the injected volumes were low (1.8 to 8.4 mL), which can lead more difficult to highlight difference. Apparent clearance (CL/F) was estimated at 0.15 L/min. This value is higher than the clearance reported for younger children during epidural block [17]. This is coherent with pharmacokinetics of levobupivacaine, metabolized by CYP3A4 and CYP1A2 isoforms. As the maturation of enzyme producing organs increases with age, an increase of mature cytochromes involve a higher clearance. We found a BSV on clearance and distribution volume similar to that reported by Chalkiadis et al. (epidural regional block, with a BSV for V and CL of 48.5% and 35.2% respectively) [17]. Our results confirm that weight must be considered when determining the dose in children.

The dosing regimen of 0.4 mg/kg is safe, no signs of toxicity were observed during the study. But an optimization can be done in terms of efficacy: TAP block effectiveness (15 min post injection) was inadequate for 13 patients out of 40. We have tried to establish a pharmacokinetic/pharmacodynamic (PKPD) model but PK parameters and efficacy were not correlated (Emax, Hill or Logit function). This finding is not surprising given that is an a posteriori analysis, the study was not designed for a PKPD analysis. The only PD endpoint was the increase of either heart rate or arterial blood pressure at the skin incision time (approximatively 15 min post injection), compared to reference values prior block placement. If the heart rate or arterial blood pressure increased, the TAP block was considered as failed (DV = 0). If they were stable, the TAP block was considered as successful (DV = 1). A PKPD model could not be establish without several and accurate PD endpoints available. We observed a significant difference in response to anaesthesia with a higher rate of success for girls (91.7%) than for boys (57.1%). One possible explanation for this difference could be related to weight. On average, girls had slightly higher weight than boys (15.7 kg for girls *versus* 14.6 kg for boys) and therefore received a higher dose into the deposit site (dose = 0.4 mg/kg). From the pharmacokinetic profiles, the plasma concentrations observed (Cmax = 0.315 mg/L and Cu,max = 3.15 µg/L, respectively) are lower than concentrations considered to be well tolerated for adults (2-4 mg/L and 110-300 µg/L, respectively [18]), even for an outlier patient (Cmax = 1.29 mg/L) (Figure 3). Considering these results, an adjustment of dosage could be considered for children, in order to improve efficiency of analgesia.

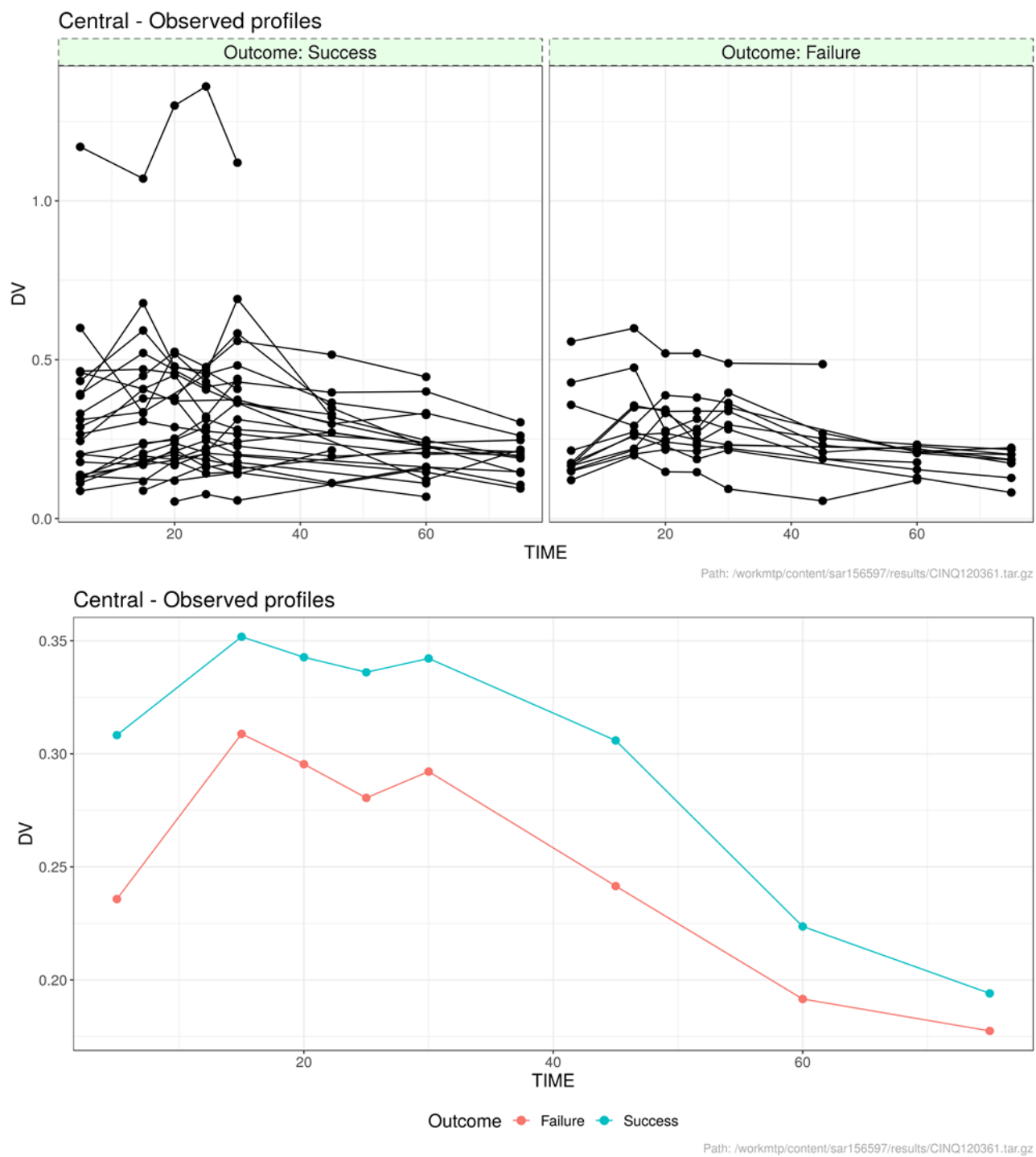

**Fig. 3** (A) Individual concentration profiles of successful TAP block, (B) Individual concentration profiles of failed TAP block, (C) Profile of average concentrations for successful TAP block (blue line) or for failed TAP block (pink line).

**5. Conclusion**

The optimal dosing regimen of levobupivacaine for an effective TAP block has not been reported in the literature and is still explored [19]. In paediatrics, even though recommendations exist, the optimal dose to be administered is not clearly established and is still empirical. At 0.4 mg/kg, none of the 1 to 5 years-old children presented toxicity signs and there is some clues that this dose could be increased, to obtain a better efficacy. Finally children's weight must be considered to evaluate their clearance and thus to anticipate any risk of toxicity.

**Acknowledgments**

Thank you to Gregory Nickson**,** for the linguistic revision.

**This research did not receive any specific grant from funding agencies in the public, commercial, or not-for-profit sectors.**